\documentclass[runningheads]{llncs}
\usepackage{soul}
\usepackage{url}
\usepackage[hidelinks]{hyperref}
\usepackage[utf8]{inputenc}
\usepackage[small]{caption}
\usepackage{graphicx}
\usepackage{proof}
\usepackage{amsmath}
\usepackage{color}
\usepackage{algorithm}
\usepackage{algorithmic}
\usepackage[labelsep=period,font=small,labelfont=bf]{caption}
\usepackage{amssymb,textcomp}
\usepackage{multirow}
\usepackage[makeroom]{cancel}
\usepackage{mathtools}
\usepackage{multicol}

\newcommand{\XG}{XG-ext}
\newcommand{\XGt}{XG-ext }

	\newcommand{\F}{\mathbb{F}}

	\title{Parity (XOR) Reasoning for the Index Calculus Attack \thanks{The final authenticated version is available online at \href{https://doi.org/10.1007/978-3-030-58475-7_45}{https://doi.org/10.1007/978-3-030-58475-7\_45}}}
	\titlerunning{Parity (XOR) Reasoning for Index Calculus}

\author{Monika Trimoska and Sorina Ionica and Gilles Dequen}

\authorrunning{M. Trimoska et al.}
\institute{Laboratoire MIS, Université de Picardie Jules Verne, Amiens, France \thanks{This work is co-financed by the European Union under the 2014/2020 European Regional Development Fund (FEDER).} \email{\{monika.trimoska,sorina.ionica,gilles.dequen\}@u-picardie.fr}}

\begin{document}

\maketitle

\begin{abstract}
Cryptographic problems can often be reduced to solving Boo- lean polynomial systems, whose equivalent logical formulas can be treated using SAT solvers. Given the algebraic nature of the problem, the use of the logical XOR operator is common in SAT-based cryptanalysis. Recent works have focused on advanced techniques for handling parity (XOR) constraints, such as the Gaussian Elimination technique. First, we propose an original XOR-reasoning SAT solver, named WDSat (Weil Descent SAT solving), dedicated to a specific cryptographic problem. Secondly, we show that in some cases Gaussian Elimination on SAT instances does not work as well as Gaussian Elimination on algebraic systems. We demonstrate how this oversight is fixed in our solver, which is adapted to read instances in algebraic normal form (ANF).
  Finally, we propose a novel preprocessing technique based on the Minimal Vertex Cover Problem in graph theory. This preprocessing technique is, within the framework of multivariate Boolean polynomial systems, used as a DLL branching selection rule that leads to quick linearization of the underlying algebraic system. Our benchmarks use a model obtained from cryptographic instances for which a significant speedup is achieved using the findings in this paper. We further explain how our preprocessing technique can be used as an assessment of the security of a cryptographic system. 
\end{abstract}

\section{Introduction} 
Cryptanalysis is the study of methods to decrypt a ciphertext without any knowledge of the secret key. Academic research in cryptanalysis is focused on deciding whether a cryptosystem is secure enough to be used in the real world. In addition, a good understanding of the complexity of a cryptographic attack allows us to determine the secret key length, making sure that no cryptanalytic effort can find the key in a feasible amount of time. Recommendations for minimum key length requirements given by various academic and governmental organizations~\cite{keylength} are based on the complexity of known attacks. 

In recent years, constraint programming (CP) techniques have been used in the cryptanalysis of both public and secret key cryptosystems. A first example in the field of differential cryptanalysis is given by the work of Gerault~\textit{et al}.~\cite{Minier_Solnon_1,Minier_Solnon_2,Minier_solnon_3} who showed how to use CP for solving the optimal related-key differential characteristic problem. Using the CP model presented in their work, all optimal related-key differential characteristics for AES-128, AES-192 and AES-256 can be computed in a few hours~\cite{Minier_solnon_3}. We also note the work of Lui~\textit{et al}.~\cite{CP-AES_1,CP-AES_2}, in which a CP model is used to aid the Tolerant Algebraic Side-Channel Analysis, which is a combination of algebraic and side-channel analysis. 

In a second line of research, Boolean satisfiability (SAT) solvers have found use in algebraic cryptanalysis. Algebraic cryptanalysis denotes any technique which reduces a cryptographic attack to the problem of solving a multivariate Boolean polynomial system. A common approach for solving these systems is to use Gr\"obner basis algorithms~\cite{Faugere99}, exhaustive search~\cite{FES:cryptoeprint:2010:313} or hybrid methods~\cite{hybrid:DBLP:journals/jmc/BettaleFP09}. These methods have been compared against SAT solving techniques for attacks on various symmetric  cryptosystems such as Bivium, Trivium, Grain. Recent work has also focused on combining algebraic and SAT solving techniques~\cite{DBLP:conf/date/ChooSCM19}. In public-key cryptography, SAT solvers have been considered for attacking binary elliptic curve cryptosystems using the index calculus attack~\cite{Galbraith14}. In this paper, we tackle this last-mentioned application.

We propose a built-from-scratch SAT solver dedicated to solving an important step of the index calculus attack. The solver, named WDSat, is adapted for XOR-reasoning and reads formulas in ANF form. In addition, we show certain limitations
of the Gaussian Elimination (GE) technique in XOR-enabled SAT solvers
  by pointing out a canceling property that is present in algebraic resolution methods but is overseen in current SAT-based GE implementations.
 We refer to this canceling property as the \XGt method and we show how it is implemented in our solver. In implementations, the \XGt method comes at a high computational cost and is thus useful only for benchmarks where it reduces significantly the number of conflicts.
Finally, we introduce a graph theory-based preprocessing technique, specifically designed for multivariate Boolean polynomial
systems, that allows us to further accelerate the resolution of our benchmarks. This preprocessing technique is designed to allow a rapid linearization of the underlying algebraic system and should be used coupled with the \XGt method. In fact, when the \XGt method is not applied, the positive outcome of the preprocessing technique cannot be guaranteed. To confirm, we perform experiments using CryptoMiniSat~\cite{Soos1:2009} coupled with our preprocessing technique and show that this combination yields slower running times than CryptoMiniSat alone. 
Experimental results in Section~\ref{sec:bench} show that the solver presented in this paper outperforms all existing solving approaches for the introduced problem. These approaches include Gröbner basis techniques~\cite{Faugere99} and state-of-the-art SAT solvers: MiniSat~\cite{minisat-first}, Glucose~\cite{glucose}, MapleLCMDistChronoBT~\cite{maple-chronoBT}, CaDiCaL~\cite{cadical-site} and CryptoMiniSat~\cite{Soos1:2009}.

\section{Background}
\subsubsection{Index Calculus}\label{sec:ic}
In cryptanalysis, the index calculus algorithm is a well-known method for attacking factoring and elliptic curve discrete logarithms, two computational problems which are at the heart of most used public-key cryptosystems. When performing this attack for elliptic curve discrete logarithms, a crucial step is the point decomposition phase. As proposed by Gaudry~\cite{Gaudry09} and Diem~\cite{Diem11} independently, a point on the elliptic curve can be decomposed into $m$ other points by solving Semaev's $(m+1)$-{th} summation polynomial~\cite{Semaev04}, that we denote by $S_{m+1}$. 
For elliptic curves defined over binary fields, the second and the third summation polynomials are defined as follows:
\begin{align}
&S_2(X_1,X_2)=X_1+X_2,& \\
&S_3(X_1,X_2,X_3)=X_1^2X_2^2+X_1^2X_3^2+X_1X_2X_3+X_2^2X_3^2+1.& \nonumber
\end{align}
For $m > 3$, the $m$-th summation polynomial is computed by using the following recursive formula:
\begin{align}
&S_m(X_1,\ldots,X_m)=& \\
&Res_X(S_{m-k}(X_1,\ldots, X_{m-k-1},X),S_{k+2}(X_{m-k},\ldots, X_m,X)),& \nonumber
\end{align}
where $Res_X$ denotes the resultant of two polynomials with respect to the $X$ variable and $1\leq k\leq m-3$.
The zeros of this polynomial will give the $x$-coordinates of points on the elliptic curve as elements in $\mathbb{F}_{2^n}$. From an implementation point of view, these will be represented as $n$-bit vectors. In index calculus attacks, the common approach is to decompose a random point given by an $n$-bit vector $x$-coordinate into $m$ points whose $x$-coordinates write as $l$-bit vectors, with $l\sim\frac{n}{m}$ (see for instance~\cite{Faugere12,Petit12}). With this choice of parameters, the problem of decomposing a random point by finding the zeros of $S_{m+1}$ can be reduced to solving a system of $n$ Boolean polynomials with $ml$ variables. 

We recall that a multivariate Boolean polynomial system is a system of polynomials in several variables and whose coefficients are in $\mathbb{F}_2$ (see for instance~\cite{LidlNiderreiter}). The following example shows a Boolean polynomial system of three equations in the variables $\{\textbf{x}_1,\textbf{x}_2,\textbf{x}_3\}$:
\begin{align*}
&\textbf{x}_1+\textbf{x}_2\cdot\textbf{x}_3 = 0 \\
&\textbf{x}_1\cdot\textbf{x}_2+\textbf{x}_2+\textbf{x}_3 = 0 \\
&\textbf{x}_1+\textbf{x}_1\cdot\textbf{x}_2\cdot\textbf{x}_3+\textbf{x}_2\cdot\textbf{x}_3 = 0.
\end{align*}
In the literature, the modelisation process allowing to obtain a Boolean polynomial system from a polynomial with coefficients in $\mathbb{F}_{2^n}$ (here the summation polynomial) is called a \textit{Weil Restriction}~\cite{Gaudry09} or \textit{Weil Descent}~\cite{Petit12}. The polynomial systems obtained in this way serve as our starting point for deriving SAT instances.\footnote{Our C code for generating these instances is publicly available~\cite{GitHubWeilDescent}.}

\subsubsection{XOR-Enabled SAT Solvers}\label{XOREnabled}

A Boolean polynomial system can be rewritten as a conjunction of logical formulas in algebraic normal form (ANF) as follows: multiplication in $\mathbb{F}_2$ ($\cdot$) becomes the logical AND operation ($\land$) and addition in $\mathbb{F}_2$ (+) becomes the logical XOR ($\oplus$). The elements 0 and 1 in $\F_2$ correspond to $\bot$ and $\top$, respectively. Consequently, solving a multivariate Boolean polynomial system is equivalent to solving a conjunction of logical formulas in ANF form. To date, few SAT solvers are adapted to tackle formulas in ANF. A common approach is to transform the ANF form in a CNF-XOR form, which is a conjunction of CNF and XOR clauses. In order to do this, every conjunction of two or more literals $x_1 \land x_2 \land \ldots \land x_k$ 
has to be replaced by an additional and equivalent variable $x'$ such that $x' \Leftrightarrow x_1 \land x_2 \land \ldots \land x_k$. This equivalence can be rewritten in CNF using a three-step transformation. First, the equivalence is decomposed into two implications:
\begin{align*}
& (x' \Rightarrow x_1 \land x_2 \land \ldots \land x_k) \; \land \\
& (x_1 \land x_2 \land \ldots \land x_k \Rightarrow x').
\end{align*}
Then, the material implication rule is applied:
\begin{align*}
& (\lnot x' \lor (x_1 \land x_2 \land \ldots \land x_k)) \; \land \\
& (\lnot(x_1 \land x_2 \land \ldots \land x_k) \lor x').
\end{align*}
Finally, using distribution on the first, and De Morgan's law on the second constraint, we obtain the following CNF formula:
\begin{align}
&(\lnot x' \lor x_1) \; \land \nonumber \\
&(\lnot x' \lor x_2) \; \land \\ 
&\ldots \nonumber \\ 
&(\lnot x' \lor x_k) \;  \land \nonumber \\
&(\lnot x_1 \lor \lnot x_2 \lor \ldots \lor \lnot x_k \lor x'). \nonumber
\end{align}
When we substitute all occurrences of conjunctions in an XOR clause by an additional variable, we obtain a formula in CNF-XOR form. This is the form used in the CryptoMiniSat solver~\cite{Soos1:2009}, which is an extension of the  MiniSat solver~\cite{minisat-first} specifically designed to work on cryptographic problems.
\begin{example}\label{transform-example}
Let us consider the Boolean polynomial system:
\begin{align}\label{example-bool-sys}
&\textbf{x}_1 + \textbf{x}_2\cdot\textbf{x}_3 + \textbf{x}_5 + \textbf{x}_6 + 1  =  0 \\
&\textbf{x}_3 + \textbf{x}_5 + \textbf{x}_6  =  0. \nonumber
\end{align}
One additional variable $x'$ needs to be introduced to substitute the monomial $\textbf{x}_2\cdot\textbf{x}_3$. The corresponding CNF-XOR form for this Boolean system is a conjunction of the following clauses:
\begin{align}\label{example-SAT}
&x' \lor \lnot x_2 \lor \lnot x_3 \nonumber \\
&\lnot x' \lor x_2 \nonumber \\
&\lnot x' \lor x_3 \\
&x_1 \oplus x' \oplus x_5 \oplus x_6 \nonumber \\
&x_3 \oplus x_5 \oplus x_6 \oplus \top. \nonumber
\end{align}
\end{example}

Finally, one could, of course, consider generic solvers (i.e. MiniSat~\cite{minisat-first}, Glucose~\cite{glucose}) for solving cryptographic problems, but this approach needs to further transform the CNF-XOR model to a CNF one. 
Transforming an XOR-clause with $k$ literals in CNF representation is a well-known process that gives $2^{k-1}$ OR-clauses of $k$ literals. 

\paragraph{Notation.} For simplicity, in the remainder of this paper we will omit the multiplication operator $\cdot$ whenever its use  in monomials is implicit. Moreover, due to equivalence between a Boolean polynomial systems and an ANF form, these will be used interchangeably. 

\section{The WDSat solver}\label{WDSat-desc}
Our WDSat solver is based on the Davis-Putnam-Logemann-Loveland (DPLL) algorithm~\cite{DBLP:journals/cacm/DavisLL62}, which is a state-of-the-art complete SAT solving technique. The solver is designed to treat ANF formulae derived from the Weil Descent modelisation of cryptographic attacks, hence its name: WDSat. The code for the WDSat solver is written in C and is publicly available~\cite{WDSat-code}.

WDSat implements three reasoning modules. These include the module for reasoning on the CNF part of the formula and the so-called XORSET and XORGAUSS (XG) modules designed for reasoning on XOR constraints. The CNF module is designed to perform classic unit propagation on OR-clauses. The XORSET module performs the operation equivalent to unit propagation, but adapted for XOR-clauses.
Practically, this consists in checking the parity of the current interpretation and propagating the unassigned literal. Finally, the XG module is designed to perform GE on the XOR constraints dynamically. We also implement an XG extension, described in Section 4. The following is a detailed explanation of this module.

XOR clauses are \text{normalized} and represented as equivalence classes. Recall that an XOR-clause is said to be in \textit{normal form} if it contains only positive literals and does not contain more than one occurrence of each literal. 
Since we consider that all variables in a clause belong to the same equivalence class (EC), we choose one literal from the EC to be the \textit{representative}. An XOR-clause $(x_1 \oplus x_2 \oplus ... \oplus x_n)~\Leftrightarrow~\top$ rewrites as 
\begin{align}\label{ec-form}
x_1~\Leftrightarrow~(x_2 \oplus x_3 \oplus ... \oplus x_n \oplus \top).
\end{align}
Finally, we replace all occurrences of a representative of an XOR clause with the right side of the equivalence. Applying this transformation, we obtain a simplified system having the following property: a representative of an EC will never be present in another EC. 

Let $R$ be the set of representatives and $C$ be the set of clauses. $R$ and $C$ hold the right-hand side and the left-hand side of all equations of type~\eqref{ec-form} respectively. We denote by $C_{x}$ the clause in $C$ that is equivalent to $x$. In other words, $C_{x}$ is the right-hand side of the EC that has $x$ as representative. Finally, we denote by $var(C_x)$ the set of literals (plus a $\top$/$\bot$ constant) in the clause $C_x$ and
$C[x_1/x_2]$ denotes the following substitution of clauses: for all $C_i \in C$ containing $x_1$, $C_i \leftarrow C_i \oplus x_1 \oplus x_2$, i.e. $x_1$ is replaced by $x_2$ in $C_i$. When we replace a literal $x_1$ by a clause $C_{x_2}$, we adopt a similar notation: $C[x_1/C_{x_2}]$.

Thus, assigning a literal $x_1$ to $\top$ leads to using one of the rules in Table~\ref{xorgauss-rules}, depending on whether $x_1$ belongs to $R$ or not. In both cases, propagation occurs when~: $\exists~x_i \neq x_1~s.t.~ var(C_{x_i}) = \top/\bot$. Conflict occurs when one constraint leads to the propagation of $x_i$ to $\top$ and another constraint leads to the propagation of $x_i$ to $\bot$.

Table~\ref{xorgauss-rules} presents inference rules for performing GE in the XG module of WDSat. 
Applying these rules allows us to maintain the property of the system which states that a representative of an EC will never be present in another EC. For clarity of the notation, the first column of this table contains the premises, the second one contains the conclusion and the third one is an update on the set $R$ which has to be performed when the inference rule is used. 

\setlength\tabcolsep{0.2cm}
\begin{table}
	\caption{Gaussian elimination inference rules.}\label{xorgauss-rules}
	\centering
	\begin{tabular}{|l|l|l|}  
	\hline
	Premises & Conclusions on $C$ & Updates on $R$ \\
	\hline
	$x_1$, $C$ & \multirow{2}{*}{$C[x_1/\top]$} &\multirow{2}{*}{\tiny{$N/A$}}\\
	$x_1 \cancel{\in} R$ & & \\
	\hline
	$x_1$, $C$  & $C_{x_2} \leftarrow C_{x_1} \oplus x_2 \oplus \top$  & $R\leftarrow R \setminus \{ x_1 \}$ \\
	$x_1\in R$ &  \multirow{2}{*}{$C[x_2/C_{x_2}]$} & \multirow{2}{*}{$R\leftarrow R \cup \{ x_2 \}$}\\
	$x_2\in var(C_{x_1})$ &&\\
	\hline
	\end{tabular}
\end{table}

We denote by $k$ the number of variables in a XOR-CNF formula. At the implementation level, XOR-clauses are represented as $(k+1)$-bit vectors: a bit for every variable and one for a $\top$,~$\bot$ constant. Clauses are stored in an array indexed by the representatives. This representation allows us to perform GE only by XOR-ing bit-vectors and flipping the clause constant. For a compact representation of the $(k+1)$-bit vector we used an array of $\lceil (k+1)/64\rceil$ integers.
\begin{example}
Let $k=7$ and let us consider $x_2~\Leftrightarrow~\top \oplus x_1 \oplus x_3 \oplus x_5$. Then we have that $var(C_{x_2})=\{\top,x_1,x_3,x_5\}$ and the bit-vector representing this clause is 11010100, where the $\top$,~$\bot$ constant takes the zero position. Assigning $x_1$ to $\top$ is equivalent to introducing the constraint $x_1 \oplus \top$. We apply the first rule, simply by XOR-ing this bit-vector with a mask of the form 11000000. The resulting vector is 00010100, which corresponds to $var(C_{x_2})=\{\bot,x_3,x_5\}$. 
\end{example}

Our DPLL-based solver assigns a truth value to each variable in a formula $F$, recursively building a binary search tree. After each assignment, either the formula is simplified and other truth values are inferred or a conflict occurs. In the case of a conflict, the last assignment has to be undone for each module via a backtracking procedure. 
In Algorithm~\ref{alg:infer}, we detail the \textsc{assign} function of WDSat, which is at the core of the DPLL algorithm. This function synchronises all three modules in the following manner. First, the truth value is assigned in the CNF module and truth values of other variables are propagated. Next, the truth value of the initial variable, as well as the propagated ones are assigned in the XORSET module. If the XOR-adapted unit propagation discovers new truth values, they are assigned in the CNF module, going back to step one. We go back and forth with this process until the two modules are synchronized and there are no more propagations left. Finally, the list of all inferred literals is transferred to the XG module. If the XG module finds new XOR-implied literals, the list is sent to the CNF module and
the process is restarted. If a conflict occurs in any of the reasoning modules, the \textsc{assign} function fails and a backtracking procedure is launched. We briefly detail the other functions used in the pseudo-code.
There is a \textsc{set\_in} function for each module which takes as input a list of literals and a propositional formula $F$ and sets all literals in this list to $\top$ in the corresponding modules. Through this assignment, the function also infers truth values of other literals, according to the specific rules in different modules. For instance, the \textsc{set\_in} function for the XG module (\textsc{set\_in\_XG}) implements the rules in Table~\ref{xorgauss-rules}, performing a GE on the system.
Finally, the \textsc{last\_assigned} function in each module returns the list of literals that were assigned during the last call to the respective \textsc{set\_in} function.

\begin{algorithm}[h!]
	\caption{Function \textsc{assign}($F$, $x$)~: 
				Assigning a truth value to a literal $x$ in a formula $F$, simplifying $F$ and inferring truth values for other literals.}
	\label{alg:infer}
	\textbf{Input}: The propositional formula $F$, a literal $x$\\
	\textbf{Output}: $\bot$ if a conflict is reached, $\top$ and a simplified $F$ otherwise
	\begin{algorithmic}[1] 
		\STATE $to\_set \leftarrow \{x\}$.
				\STATE $to\_set\_in\_XG \leftarrow \{x\}$.
				\WHILE{$to\_set \neq \emptyset$}
				\WHILE{$to\_set \neq \emptyset$}
				\IF {\textsc{set\_in\_CNF}($to\_set$, $F$) $\rightarrow \bot$}
				\STATE \textbf{return} ($\bot$, -- ).
				\ENDIF
				\STATE $to\_set \leftarrow$ \textsc{last\_assigned\_in\_CNF()}.
				\STATE $to\_set\_in\_XG \leftarrow to\_set$.
				\IF {\textsc{set\_in\_XORSET}($to\_set$, $F$)$\rightarrow \bot$}
				\STATE \textbf{return} ($\bot$, -- ).
				\ENDIF
				\STATE $to\_set \leftarrow$ \textsc{last\_assigned\_in\_XORSET()}.
				\STATE $to\_set\_in\_XG \leftarrow to\_set \cup to\_set\_in\_XG$.
				\ENDWHILE
				\IF {\textsc{set\_in\_XG}($to\_set\_in\_XG$, $F$)$\rightarrow \bot$}
				\STATE  \textbf{return} ($\bot$, -- ).
				\ENDIF
				\STATE $to\_set \leftarrow$ \textsc{last\_assigned\_XG}().
				\ENDWHILE
				\STATE \textbf{return} ($\top$, $F$).
	\end{algorithmic}
\end{algorithm}

\section{The XG-ext Method}\label{XG-method}
In this section, we show how we extend our XG module. First, we present the motivation for this work by giving an example of a case where GE in SAT solvers has certain limitations compared to Algebraic GE. Secondly, we propose a solution to overcome these limitations and we implement it in our solver to develop the XORGAUSS-ext method (\XGt in short). To introduce new rules for this method, we use the same notation as in Section~\ref{WDSat-desc}.

Gaussian elimination on a Boolean polynomial system consists in performing elementary operations on equations with the goal of reducing the number of equations as well as the number of terms in each equation. We cancel out terms by adding (XOR-ing) one equation to another. GE can be performed on instances in CNF-XOR form in the same way that it is performed on Boolean polynomial systems presented in algebraic writing. 
However, we detected a case where a possible cancellation of terms is overseen due to the CNF-XOR form. 
\begin{example}\label{cancellation}
We will reuse the Boolean polynomial system in Example~\ref{transform-example} to demonstrate a case where a cancellation of a term is missed by a XOR-enabled SAT solver.
Let us consider that in Equation~\eqref{example-bool-sys}, we try to assign the value of 1 to $\textbf{x}_2$. As the monomial $\textbf{x}_2\textbf{x}_3$ will be equal to 1 only if both terms $\textbf{x}_2$ and $\textbf{x}_3$ are equal to 1, we get the following result:
\begin{align*}
&\textbf{x}_1 + \textbf{x}_3 + \textbf{x}_5 + \textbf{x}_6 + 1  =  0 \\
&\textbf{x}_3 + \textbf{x}_5 + \textbf{x}_6  =  0.
\end{align*}
After XOR-ing the two equations, we infer that  $\textbf{x}_1 = 1$. 

\noindent
However, when we assign $x_2$ to $\top$ in the corresponding CNF-XOR clause in Equation~\eqref{example-SAT}, as per unit propagation rules, we get the following result:
\begin{align*}
&x' \lor \lnot x_3 \\
&\lnot x' \lor x_3 \\
&x_1 \oplus x' \oplus x_5 \oplus x_6 \\
&x_3 \oplus x_5 \oplus x_6 \oplus \top.
\end{align*}
When we XOR the second clause to the first one we can not infer that $x_1$ is $\top$ at this point.

Note that $(x' \lor \lnot x_3) \land (\lnot x' \lor x_3)$ rewrites as $x' \Leftrightarrow x_3$, but if the solver does not syntactically search for this type of occurrences regularly, $x'$ will not be replaced by $x_3$. Moreover, this type of search adds an additional computational cost to the resolution.
\end{example}

Omissions as the one detailed in Example~\ref{cancellation} can occur every time a variable is set to $\top$. As a result, we define the following rule with the goal to improve the performance of XOR-enabled SAT solvers:
\begin{align}\label{subst-rule}
\infer{x_1\Leftrightarrow x_2}{x' & x_1 \Leftrightarrow (x' \land x_2) }.
\end{align}
This rule can be generalised for the resolution of higher-degree Boolean polynomial systems:
\begin{align}\label{subst-rule-gen}
\infer{x_1\Leftrightarrow (x_2 \land \ldots \land x_d)}{x' & x_1 \Leftrightarrow (x' \land x_2 \land \ldots \land x_d) }.
\end{align}
Even though these rules are standard in Boolean logic, they are presently not implemented in XOR-enabled SAT solvers. Note that when a solver takes as input an instance in CNF-XOR form, the second premise is lost or has to be inferred by syntactic search. To have knowledge of the second premise, the solver needs to read the instance in ANF. To this purpose, we defined a new ANF input format for SAT solvers.

\begin{table}[h]
	\caption{Inference rules for the substitution of $x_1$ by $x_2$.}\label{xorgauss-en-rules}
	\centering
	\begin{tabular}{|l|l|l|}  
		\hline
		Premises & Conclusions on $C$ & Updates on $R$ \\
		\hline
		\small{$C$, $x_1\Leftrightarrow x_2$}	 & \multirow{3}{*}{\small{$C[x_1/x_2]$}}& \multirow{3}{*}{\small{$N/A$}}\\
		\small{$x_1 \cancel{\in} R$}& &\\
		\small{$x_2 \cancel{\in} R$}& & \\
		\hline
		\small{$C$, $x_1\Leftrightarrow x_2$} &  \multirow{2}{*}{\small{$C_{x_2} \leftarrow C_{x_1}$}} & \multirow{2}{*}{\small{$R\leftarrow R \setminus \{ x_1 \}$}}\\ 
		\small{$x_1 \in R$} & &\\ \small{$x_2 \cancel{\in} R$} & \multirow{2}{*}{\small{$C[x_2/C_{x_2}]$}} & \multirow{2}{*}{\small{$R\leftarrow R \cup \{ x_2 \}$}}\\
		\small{$x_2 \cancel{\in}  var(C_{x_1})$}&   &  \\
		\hline
		\small{$C$, $x_1\Leftrightarrow x_2$} & \multirow{2}{*}{\small{$C_{x_3} \leftarrow C_{x_1} \oplus x_2 \oplus x_3$}}
		& \multirow{2}{*}{\small{$R\leftarrow R \setminus \{ x_1 \}$}} \\ 
		\small{$x_1 \in R$}   & & \\
		\small{$x_2 \cancel{\in} R$} & \multirow{3}{*}{\small{$C[x_3/C_{x_3}]$}} & \multirow{3}{*}{\small{$R\leftarrow R \cup \{ x_3 \}$}} \\
		\small{$x_2 \in  var(C_{x_1})$}  &  &  \\
		\small{$x_3 \in  var(C_{x_1})$} & 	 & \\
		\hline
		\small{$C$, $x_1\Leftrightarrow x_2$}	& 	\multirow{4}{*}{\small{$C[x_1/C_{x_2}]$}} &	\multirow{4}{*}{\small{$N/A$}}\\
		\small{$x_1 \cancel{\in} R$} & & \\
		\small{$x_2 \in R$} & & \\
		\small{$x_1 \cancel{\in} var(C_{x_2})$} & &\\
		\hline
		\small{$C$, $x_1\Leftrightarrow x_2$}	 & \multirow{2}{*}{\small{$C_{x_3} \leftarrow C_{x_2} \oplus x_1 \oplus x_3$}} & \multirow{2}{*}{\small{$R\leftarrow R \setminus \{ x_2 \}$}}   \\
		\small{$x_1 \cancel{\in} R$}   &   & \\
		\small{$x_2 \in R$}  & \multirow{3}{*}{\small{$C[x_1/x_2,x_3/C_{x_3}]$}}& 	\multirow{3}{*}{\small{$R\leftarrow R \cup \{ x_3 \}$}} \\
		\small{$x_1 \in  var(C_{x_2})$}	& &\\
		\small{$x_3 \in  var(C_{x_2})$}	&   &  \\
		\hline
		\small{$C$, $x_1\Leftrightarrow x_2$} 	 &  \multirow{2}{*}{\small{$C_{x_3} \leftarrow C_{x_1} \oplus C_{x_2} \oplus x_3$}} & \multirow{2}{*}{\small{$R\leftarrow R \setminus \{x_1, x_2 \}$}} \\ 
		\small{$x_1 \in R$} & & \\
		\small{$x_2 \in R$} & \multirow{2}{*}{$C[x_3/C_{x_3}]$} & \multirow{2}{*}{\small{$R\leftarrow R \cup \{ x_3 \}$}} \\
		\small{$x_3 \in var(C_{x_1}\oplus C_{x_2})$} & & \\
		\hline
	\end{tabular}
\end{table}

This extension of the XG module is implemented as part of the \textsc{set\_in\_XG} function used in the \textsc{assign} algorithm.
The following is a detailed explanation of how the rule in Equation~\eqref{subst-rule} is applied in our implementation. Recall that the XG module has the following property: a representative of an EC will never be present in another EC. This property will be maintained in the \XGt method as well. Using the conclusion in Equation~\eqref{subst-rule}, we derive in Table~\ref{xorgauss-en-rules} six inference rules that allow us to perform the substitution of a variable $x_1$ by a variable $x_2$ while maintaining the unicity-of-representatives property. Applying one of the inference rules in Table~\ref{xorgauss-en-rules} can result in conflict or it can propagate a newly discovered truth value. Note that $var(C_{x_1}\oplus C_{x_2})$ is given by the symmetric difference $(var(C_{x_1}) \cup var(C_{x_2})) \setminus (var(C_{x_2}) \cap var(C_{x_1}))$.

	\section{Our Preprocessing Technique}
	Let us reconsider the DPLL-based algorithm. It is well known that the number of conflicts needed to prove the inconsistency is correlated to the order in which the variables are assigned. Among the state-of-the-art branching rules you can find two categories according to the type of heuristics. The first are based on Maximum number of Occurrences in the Minimum clauses Size (MOMs) whereas the second adopt the Variable State Independent Decaying Sum (VSIDS) branching heuristic.
	
	In this work, we were interested in developing a criterion for defining the order of variables on CNF-XOR instances derived from Boolean polynomial systems. We set the goal to choose branching variables that will lead as fast as possible to a linear polynomial system, which can be solved using GE in polynomial time. In terms of SAT solving, choosing this order for branching will cancel out all clauses in the CNF part of the formula as a result of unit propagation. When only the XOR part of the CNF-XOR formula is left, the solver performs GE on the remaining XOR constraints in polynomial time.  
	
	After setting this goal, choosing which variable to assign next according to the number of their occurrences in the system is no longer an optimal technique. We explain this idea on an example. For simplicity, we only use the Boolean algebra terminology in this section. However, the methods described are applicable to both SAT solving and algebraic techniques based on the process of recursively making assumptions on the truth values of variables in the system (as with the DPLL algorithm).
	
	\begin{example}\label{MVC-example-sys}
	Consider the following Boolean polynomial system:
	\begin{align}
	&\textbf{x}_1 + \textbf{x}_2\textbf{x}_3 + \textbf{x}_4 + \textbf{x}_4\textbf{x}_5 = 0 \\ \nonumber
	&\textbf{x}_1 + \textbf{x}_2\textbf{x}_3 = 0 \\ \nonumber
	&\textbf{x}_1 + \textbf{x}_3\textbf{x}_5 + \textbf{x}_6 = 0 \\ \nonumber
	&\textbf{x}_1 + \textbf{x}_2\textbf{x}_5\textbf{x}_6+ \textbf{x}_6 = 0 \nonumber
	\end{align}
	In this example, the variable with the highest number of occurrences is $\textbf{x}_1$. However, $\textbf{x}_1$ does not occur in any monomial of degree $> 1$. Thus, assigning first $\textbf{x}_1$ does not contribute to the linearization of the system and we need to find a more suitable criterion.
	\end{example}
	
	The solution we propose is inspired by graph theory. Particularly, we identified a parallel between the problem of defining the order in which the variables are assigned and the Minimal Vertex Cover Problem (MVC). 
	
	In graph theory, a \textit{vertex cover} is a subset of vertices such that for every edge $(v_i,v_j)$ of the graph, either $v_i$ or $v_j$ is in the vertex cover. Given an undirected graph, the Minimum Vertex Cover Problem is a classic optimization problem of finding a vertex cover of minimal size. 
	
	An undirected graph is derived from a Boolean polynomial system as follows.
\begin{itemize}
	\item Each variable $\textbf{x}_i$ from the system becomes a vertex $v_i$ in the graph $G$.
	\item An edge $(v_i,v_j)$ is in $G$ if and only if (in the corresponding Boolean system) there exists a monomial of degree $n \geq 2$  which contains both $\textbf{x}_i$ and $\textbf{x}_j$.
\end{itemize}
	
	When we use this representation of a Boolean polynomial system as a graph, a vertex cover defines a subset of variables whose assignment will result in a linear Boolean polynomial system in the remaining non-assigned variables. Consequently, finding the MVC of the graph is equivalent to finding the minimal subset of variables one has to assign to obtain a linear system. 
	
	\begin{figure}
		\centerline{\includegraphics[height=1.5in]{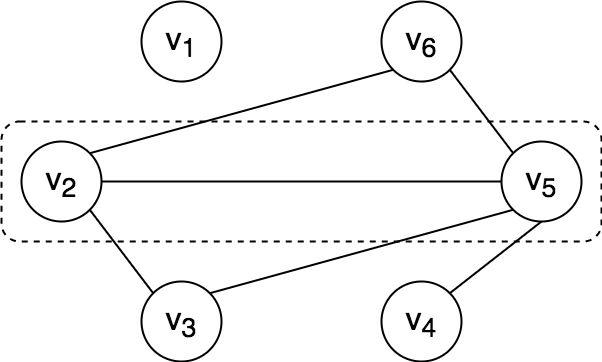}}
		\caption{Graph derived from Example~\ref{MVC-example-sys}} \label{MVC-example-graph}
	\end{figure}

	Figure~\ref{MVC-example-graph} shows the graph derived from Example~\ref{MVC-example-sys}. The MVC of this graph is $\{v_2, v_5\}$. As a result, when all variables in the subset $\{\textbf{x}_2, \textbf{x}_5\}$ are assigned, the remaining polynomial system is linear. We give here the system derived after the assignment $\textbf{x}_2 = 1$ and  $\textbf{x}_5 = 1$.
	\begin{align*}
	&\textbf{x}_1 + \textbf{x}_3 = 0 \\
	&\textbf{x}_1 + \textbf{x}_3 + \textbf{x}_6 = 0 \\
	&\textbf{x}_1 = 0.
    \end{align*}
    For all other possible assignments of $\textbf{x}_2$ and $\textbf{x}_5$, we obtain similar linear systems.
	
	Defining the order of branching variables will serve as a preprocessing technique that consists in (i) deriving a graph from a Boolean polynomial system and (ii) finding the MVC of the resulting graph. During the solving process, variables corresponding to vertices in the MVC are assigned first. Even though the MVC problem is NP-complete, its execution for graphs derived from cryptographic models always finishes in negligible running time due to the small number of variables. Our solver does not use any other MOMs or VSIDS-based heuristic during the solving process, as the order of the branching variables is predetermined by the MVC preprocessing technique.
	
	When variables are assigned in the order defined by this preprocessing technique, the worst-case time complexity of a DPLL-based algorithm drops from $O(2^k)$ to $O(2^{k'})$, where $k'$ is the number of vertices in the MVC set. Note that the MVC of a complete graph is equal to the number of its vertices. Consequently, when the corresponding graph of a Boolean polynomial system is a complete graph, solving the system using this preprocessing technique is as hard as solving the system without it. 
	
	Finding the MVC corresponding to a Boolean polynomial system can also be used as an assessment of the security of the underlying cryptosystem. Indeed, an exhaustive search on a subset of variables, which are the variables in the MVC, results in linear systems that can be solved in polynomial time. This straightforward approach yields an upper bound on the complexity of solving the system at hand. In short, to assess the security of a cryptographic system,
assuming that this is based on solving the Boolean polynomial system first, one computes the MVC
of this system and deduces that $O(2^{k'})$ is a bound on the complexity of the attack.
	
	\section{Experimental Results}\label{sec:bench}
	To support our claims, we experimented with benchmarks derived from two variants of the index calculus attack on the discrete logarithm problem over binary elliptic curves. As explained in Section~\ref{sec:ic}, a SAT solver can be used for solving Semaev's summation polynomials in the point decomposition phase. Our model is derived from the Boolean multivariate polynomial system given by the $m+1$-th summation polynomial, with $m\geq 2$. This model has previously been examined in~\cite{Galbraith14}. We compare the WDSat solver presented in this paper to the following approaches: the best currently available implementation of Gr\"obner basis (F4~\cite{Faugere99} in MAGMA~\cite{MR1484478}), the solvers MiniSat,~\cite{minisat-first}, Glucose~\cite{glucose}, MapleLCMDistChronoBT~\cite{maple-chronoBT}, CaDiCaL~\cite{cadical-site} and CryptoMiniSat~\cite{Soos1:2009} with enabled GE.\footnote{Enabling GE in CryptoMiniSat yielded better performance for these benchmarks.} Note that MapleLCMDistChronoBT and CaDiCaL are the winners in the main track of the latest SAT competition~\cite{SatCompet} in 2018. All tests were performed on a 2.40GHz Intel Xeon E5-2640 processor and are an average of 100 runs.

    For SAT models derived from cryptographic problems, the preprocessing technique is executed only once, since all instances presenting a specific cryptographic problem are equivalent except for the constant in the XOR constraints. Even though the MVC problem is NP-complete, its execution for graphs derived from our models always finished in negligible running time, due to the small number of nodes.
	
	We conducted experiments using both the third and the fourth polynomials. Results on solving the third summation polynomial ($m=2$) are shown in Table~\ref{res-third1}. The parameters used to obtain these benchmarks are $n=41$ and $l=20$. As a result, we obtained a Boolean polynomial system of 41 equations in 40 variables (see Section~\ref{sec:ic}). We show running-time averages on satisfiable and unsatisfiable instances separately, as these values differ between the two cases. 
	
	As different variants of our solver can yield better results for different benchmarks, we compared all variants to decide on the optimal one. We also tested the solver with and without our preprocessing technique (denoted by mvc in the tables). The results in Table~\ref{res-third2} show that WDSat yields optimal results for these benchmarks when the \XGt method is used coupled with the preprocessing technique. This outcome is not surprising when we examine the MVC obtained by the preprocessing technique. The number of variables in the system is $k=40$, but the number of vertices in the MVC is $20$. This means that by using the optimization techniques described in this paper, the worst-case time complexity of the examined models drops from $2^{k}$ to $2^{\frac{k}{2}}$. This is the case for every instance derived from the third summation polynomial.
	\bgroup
    \def\arraystretch{1.1}
	\begin{table}
		\caption{Comparing different versions of WDSat for solving the third summation polynomial.}\label{res-third2}
		\centering
		\begin{tabular}{|l|ll|ll|}
		    \hline
			\multirow{2}{*}{WDSat+}&\multicolumn{2}{c|}{SAT}&\multicolumn{2}{c|}{UNSAT} \\
			\cline{2-5}
			&Runtime (s)&\#Conflicts&Runtime (s)&\#Conflicts \\
			\hline
			XG&6028.4&200957178&11743.2&354094821 \\
			\hline
			XG+mvc&639.6&21865963&2973.0&94489361 \\
			\hline
			\XG&375.9&4911099&870.1&10789518 \\
			\hline
			\XG+mvc&\textbf{4.2}&\textbf{27684}&\textbf{13.5}&\textbf{86152} \\
			\hline
		\end{tabular}
	\end{table}

By analyzing the average running time and the average number of conflicts in Table~\ref{res-third1}, we see that the chosen variant of the WDSat solver outperforms all other approaches for solving instances derived from the third summation polynomial.

Current versions of CryptoMiniSat do not allow choosing the order of the branching variables as its authors claim that this technique almost always results in slower running times. To verify this claim, we modified the source code of CryptoMiniSat in order to test our preprocessing technique coupled with this solver (see line CryptoMiniSat+mvc in Table~\ref{res-third1}). We set a timeout of 10 minutes and only 9 out of 100 unsatisfiable and 54 out of 100 satisfiable instances were solved. This confirms that the  MVC preprocessing technique is strongly linked to our \XGt method. Indeed, when the \XGt method is not used, one can not guarantee that when all variables from the MVC are assigned the system becomes linear. This is confirmed also by looking at the number of conflicts for the CryptoMiniSat+mvc approach, which is greater than $2^{\frac{k}{2}}$ even for benchmarks that were solved before the timeout. Recall that $\frac{k}{2}$ is the size of the MVC. On the other hand CryptoMiniSat without the preprocessing technique succeeds in solving these instances after less than $2^{\frac{k}{2}}$ conflicts. We conclude that the searching technique in CryptoMiniSat used to decide on the next branching variable is optimal for this solver.

The solvers which are not XOR-enabled did not solve any of the 200 satisfiable and unsatisfiable instances before the 10-minute timeout. This is not surprising as instances derived from the third summation polynomial are solved a lot faster when a GE technique is used.

\begin{table}
	\caption{Comparing different approaches for solving the third summation polynomial.}\label{res-third1}
	\centering
	\begin{tabular}{|l|ll|ll|}  
		\hline
			\multirow{2}{*}{Solving approach}&\multicolumn{2}{c|}{SAT}&\multicolumn{2}{c|}{UNSAT} \\
			\cline{2-5}
			&Runtime (s)&\#Conflicts&Runtime (s)&\#Conflicts \\
			\hline
		Gr\"obner&16.8&\small{$N/A$}&18.7&\small{$N/A$} \\
		\hline
		MiniSat&$>$ 600&&$>$ 600& \\
		\hline
		Glucose&$>$ 600&&$>$ 600& \\
		\hline
		MapleLCMDistChronoBT&$>$ 600&&$>$ 600& \\
		\hline
		CaDiCaL&$>$ 600&&$>$ 600& \\
		\hline
		CryptoMiniSat&29.0&226668&84.3&627539 \\
		\hline
		CryptoMiniSat+mvc&237.4&1263601&$>$ 600& \\
		\hline
		WDSat+\XG+mvc&\textbf{4.2}&\textbf{27684}&\textbf{13.5}&\textbf{86152} \\
		\hline
	\end{tabular}
\end{table} 

Experimental results in Table~\ref{res-fourth} are performed using benchmarks derived from the fourth summation polynomial. We obtain our model using a symmetrization technique proposed by Gaudry~\cite{Gaudry09}. According to our parameter choice, the initial polynomial system contains 52 equations in 51 variables. However, only 18 out of the 51 variables are 'crucial'. The other 33 variables are introduced as a result of Gaudry's symmetrization technique. Our experiments show that performing GE on these instances does not result in faster running times. On the contrary, running times are significantly slower when the XG module of the WDSat solver is enabled. Running times become even slower with the \XGt method. We attribute this fallout to the particularly small improvement in the number of conflicts, compared to the significant computational cost of performing the GE technique. Indeed, the graph corresponding to the model for the fourth summation polynomial is complete and thus the size of the MVC is equivalent to the number of variables in the formula. This leads us to believe there is no optimal choice for the order of branching variables and the system generally does not become linear until the second-to-last branching. We conclude that for solving these instances WDSat without GE is the optimal variant, since it outperforms both the Gr\"obner basis method and current state-of-the-art solvers. 

To sum up, when WDSat is used for the index calculus attack, our recommendation is to enable the XG-ext option for instances obtained from the third summation polynomial and to completely disable the XG module for instances from the fourth polynomial. For ANF instances arising from other cryptographic problems, it would be best to solve smaller instances of the problem and analyse the number of conflicts. If the number of conflicts is only slightly better when the XG module is enabled, then disabling the XG module is likely to yield faster running times for higher scale instances of that problem.

\begin{table}
	\caption{Comparing different approaches for solving the fourth summation polynomial.}\label{res-fourth}
	\centering
	\begin{tabular}{|l|ll|ll|}  
		\hline
			\multirow{2}{*}{Solving approach}&\multicolumn{2}{c|}{SAT}&\multicolumn{2}{c|}{UNSAT} \\
			\cline{2-5}
			&Runtime (s)&\#Conflicts&Runtime (s)&\#Conflicts \\
			\hline
		Gr\"obner&229.3&\small{$N/A$}&229.4&\small{$N/A$} \\
		\hline
		MiniSat&239.7&1840190&517.0&3433304 \\
		\hline
		Glucose&189.2&1527158&274.8&2056575 \\
		\hline
		MapleLCMDistChronoBT&655.1&4035131&918.7&5378945 \\
		\hline
		CaDiCaL&43.6&254194&141.3&629869 \\
		\hline
		CryptoMiniSat&331.8&1791188&707.9&3416526 \\
		\hline
		WDSat&\textbf{0.6}&\textbf{48438}&\textbf{3.8}&\textbf{255698} \\
		\hline
		WDSat+XG&19.0&85282&49.8&252949 \\
		\hline
	\end{tabular}
\end{table}

Our solver is dedicated to problems arising from a Weil descent. However, we tested it on Trivium~\cite{Canniere05trivium-} instances as they are extensively used in the SAT literature. We created instances using a modelization similar to the one in Grain of Salt~\cite{soos:hal-01288922}, a tool for deriving instances for keystream generators comprised of Nonlinear-Feedback Shift Registers (NLFSR). Our experience is that CryptoMiniSat yields faster running times than all of the WDSat variants for Trivium instances. WDSat does not implement any of the optimizations for Trivium such as dependent variable removal, sub-problem detection, etc. as there are no such occurrences in systems arising from a Weil descent.

	\section{Conclusion}
	In this paper, we revisited XOR-enabled SAT solvers and their use in cryptanalysis. We proposed a novel SAT solver, named WDSat, dedicated to solving instances derived from the index calculus attack on binary elliptic curves.
	We conducted experiments comparing WDSat to the algebraic Gröbner basis resolution method, as well as to five state-of-the-art SAT solvers. Our solver outperforms all existing resolution approaches for this specific problem.
	
	\bibliographystyle{splncs04}
	\bibliography{biblio}
\end{document}